# A Comparative Study for Predicting Heart Diseases Using Data Mining Classification Methods

Isra'a  Ahmed Zriqat, Ahmad Mousa Altamimi, Mohammad Azzeh
Faculty of Information Technology
Applied Science Private University
Amman, Jordan
{i_zriqat, a_altamimi, m.y.azzah}@asu.edu.jo

*Abstract-* Improving the precision of heart diseases detection has been investigated by many researchers in the literature. Such improvement induced by the overwhelming health care expenditures and erroneous diagnosis. As a result, various methodologies have been proposed to analyze the disease factors aiming to decrease the physicians practice variation and reduce medical costs and errors. In this paper, our main motivation is to develop an effective intelligent medical decision support system based on data mining techniques. In this context, five data mining classifying algorithms, with large datasets, have been utilized to assess and analyze the risk factors statistically related to heart diseases in order to compare the performance of the implemented classifiers (e.g., Naïve Bayes, Decision Tree, Discriminant, Random Forest, and Support Vector Machine). To underscore the practical viability of our approach, the selected classifiers have been implemented using MATLAB tool with two datasets. Results of the conducted experiments showed that all classification algorithms are predictive and can give relatively correct answer. However, the decision tree outperforms other classifiers with an accuracy rate of 99.0% followed by Random forest. That is the case because both of them have relatively same mechanism but the Random forest can build ensemble of decision tree. Although ensemble learning has been proved to produce superior results, but in our case the decision tree has outperformed its ensemble version.

## I. INTRODUCTION

Data mining techniques have been widely used for variety of applications. In health care industry for example, data mining plays an important role for predicting or diagnosing diseases with good accuracy. One important application is to diagnose the heart diseases or cardiovascular as these diseases are recognized as the leading cause of death globally in our modern world [1]. According to the World Heart Federation and the World Health Organization, more than 17 million people died from cardiovascular diseases in 2013, and around 3 million of these deaths occurred before the age of 60 [2]. However, 90% of those deaths were estimated to be preventable if patients have correctly been diagnosed early and they improved their habits such as: healthy eating, exercise, and alike [3].

In traditional healthcare environments, diagnosis of a disease depends on doctor's decision for identifying it as the most likely cause depending on a person's symptoms. However, this leads to unwanted errors that resulting on more medical costs and affecting the quality of service provided to patients. Instead, expert systems (that use Data mining techniques) [4] could be used to emulate the decision-making ability of a human expert for answering not only simple questions like "What is the average age of patients who have heart disease?", "Identify the female patients who are single, and who have been treated for heart diseases?", but also complex ones like "Given patient records, predict the probability of patients who diagnosed a heart disease?", "Find the most significant risk factor that results a heart disease?". Off course, using such systems could reduce medical errors, and decrease practice variation, but surprisingly it can improve diagnose results.

Techniques of data mining can be used for discovering knowledge in huge volumes of data through detecting patterns and summarizing data into a format that can be understood. In fact, there are three main techniques of data mining that can be utilized to classify previously unorganized data into predefined classes, or to categorize new data into pre-existing categories. This could be done by examining the data that has previously been classified, learning the rules of classification and applying those rules to new data. Or to identify relationships between them and develop a pattern of these relationships. These patterns could be then used as a reference to predict future behavior [5]. Surveys such as [6, 7] have been discussed the impact and power of data mining techniques in predicting systems.

In this paper, we have focused on data mining classification techniques that are capable of predicting a certain outcome based on a given input. In particular, we have utilized five classifiers to analyze a medical dataset that recorded previously to diagnosis cardiovascular diseases. Number of experiments have been conducted to compare the performance of the implemented classifiers (e.g., Naïve Bayes, Decision Tree, Discriminant, Random Forest, and Support Vector Machine) on a different size full training dataset with 14 clinical attributes. Results showed that Decision tree outperforms other classifiers with an accuracy rate of 99.0%, which provided a more effective and comprehensive forecasting mechanism that can be integrated with the medical information system to assist in the diagnosis of the heart diseases in the earlier stages.

The rest of the paper is organized as follows. The reviewing of some related works to the proposed approach is presented in Section II. Basic concepts related to our research are introduced in Section III. Section IV presents various classification algorithms. Section V discusses the research methodology. The results and discussions are presented in Section VI. The final conclusions and future works are offered in Section VII.

## II. RELATED WORK

Many experiments have been carried out for diagnosing heart diseases [2, 6 – 25]. In this context, researchers have been applied different data mining techniques such as: Association Rules Technique, Clustering, and Classification Algorithms to extract significant patterns for prediction of heart diseases with good accuracy.

The association rules mining technique, for example, has been utilized in many works to find frequent item set among large patients data sets to diagnose the presence of heart diseases. Authors of [2, 8, 9] used association rules to build a model for determining relations amongst patients' attributes values for discovering interesting relations between variables in the dataset. Others [11, 12, 13] have followed the same principle with little modifications on their models. [11] for instance, optimized the association rules in case of huge generated set to save the prediction time and avoid the unwanted rules. [13] proposed a new approach based on sequence number and clustering small amount of data at a time to produce scalable and efficient predicting of heart diseases. An integrated approach of association and classification was also introduced for discovering rules in the database and then using these rules an efficient classifier is constructed [16]. However, while association algorithms have been successfully applied for prediction in health domain, they discovered huge number of rules (some of them are not interested), in addition generally the performance of association algorithms is low [23].

On the other hand, Clustering is also employed as a process of partitioning a patients set into a set of meaningful clusters. Each of which discovers overall distribution pattern and correlations among data attributes to help users understanding the natural grouping of patients' data to diagnose correctly the presence of heart diseases [26]. However, clustering becomes so expensive in case of processing large data set, so it can be used as a pre-processing step for other classification algorithms to get insight into data distribution. That being said, clustering techniques have been applied to find out hidden patterns related to heart patients. Authors of [14, 15] used K-means clustering algorithm to cluster a heart disease dataset to extract data relevant to heart disease. [15] also applied the Maximal Frequent Item set algorithm (MAFIA) for mining maximal frequent patterns significant to heart attack predictions. It is important to mention that many clustering algorithms were implemented in the literature with the help of WEKA tool [27], which is a collection of machine learning algorithms and tools used for data pre-processing, classification, clustering, and visualization. However, the main drawbacks of such methods are that they do not work well with clusters (in the original data) of different size and different density, it is difficult to predict K-Value, and in case of choosing different initial partitions different final clusters could be produced [24, 25].

Genetic Algorithm has also been used in [10], to optimize the data size to get the optimal subset amongst patients' attributes values that are sufficient for heart disease prediction. The global optimization benefit of genetic algorithm has been utilized in [21] to implement a hybrid system for initialization of neural network weights based on a set of risk factors such as hypertension, high cholesterol, obesity, etc. In another system [22], a layered neuro-fuzzy approach has been utilized to produce a very low error rate in performing analysis for heart disease occurrences.

Ultimately, classification algorithms have also been extensively applied to develop a heart diseases prediction models. Popular data mining algorithms such as: Support Vector Machine, Decision Trees, Naïve Bayes, and Neural Networks have been utilized in this context. Diverse results have been presented in [17, 18, 19, 20] when the aforementioned classification algorithms were employed to develop prediction models. For instance, authors of [17] used Support Vector Machine, Artificial Neural Network, and Decision Tree with 502 cases. Their experiments showed that Support Vector Machine was the best prediction model followed by Artificial Neural Networks. In contrast, results of [20] proved that Naïve Bayes technique outperformed the other used algorithms (Decision Tree, K-NN and Neural Network) in the domain of heart disease diagnoses, while the Neural Network was recognized as the best prediction model in [18] when Decision Trees, Naïve Bayes, and Neural Network were used with 15 popular attributes as risk factors listed in the medical literature. To improve the accuracy of prediction, the hybrid intelligent technique has been bind to neural networks in [19] to produce a prediction model. The results showed that the hybrid intelligent technique improved the accuracy of prediction.

After reviewing the classification related literatures, we were motivated to utilize the classification algorithms to develop a prediction model for heart diseases based on patterns generated from clinical different sized data sources. Our work is different from the works presented above in two ways: Firstly, five classifiers methods have been applied to analyze medical datasets with different sizes. Secondly, an extensive analysis has been conducted using MATLAB tool. The simulated implementation produced a very low error rate when calculating the prediction's accuracy.

### III. OVERVIEW OF DATA MINING TECHNIQUES

Healthcare environment generally includes rich data about patients and their treatment, which are stored in health management systems. Such data is valuable especially if we cultivate the existing information into useful practices. Data mining techniques can help in extracting valuable knowledge from hidden relationships and trends among data. In fact, data mining techniques have been extensively used in healthcare research as a stage of knowledge discovery process [1], which offers promising ways to uncover hidden patterns within large amounts of health data [3, 28]. It has been applied to a diversity of healthcare domains to improve and accelerate decision making [1]. The data mining techniques are divided into four types: Classification, Clustering, Regression and Association rule mining.

Classification methods are the most widely used algorithms in Healthcare sector as it helps predicting the status of patient by classifying patient records and find the class that matches the new patient record [3]. Basically, classification is known as supervised learning techniques which requires the data to be initially classified into initial classes or labels [29]. Then these data are entered into a classification algorithm in order to be learned as shown in Figure 1. Particularly, the relationship between attributes needs to be discovered by the algorithm to predict the outcome. In this phase the classification algorithms builds the classifier from the training set made up of dataset tuples and their associated class labels. Every tuple that constitutes the training set is referred to as a category or class. When a new case is arrived the developed classification algorithm is used to classify it into one of the predefined classes as shown in Figure 2. The term which specifies how "good the algorithm is" is called prediction accuracy. For instance, the training set in the medical database would have much relevant patient information recorded already, where the prediction outcome is whether or not the patient had a heart disease [3, 28]. In literature there are many classification algorithm that can be used for healthcare datasets. Among them, we have chosen Decisions Tree, Naive Bayes, Discriminant, Random forest, and Support vector Machine (SVM). In the next section we give an overview to each one of the employed classification algorithms.

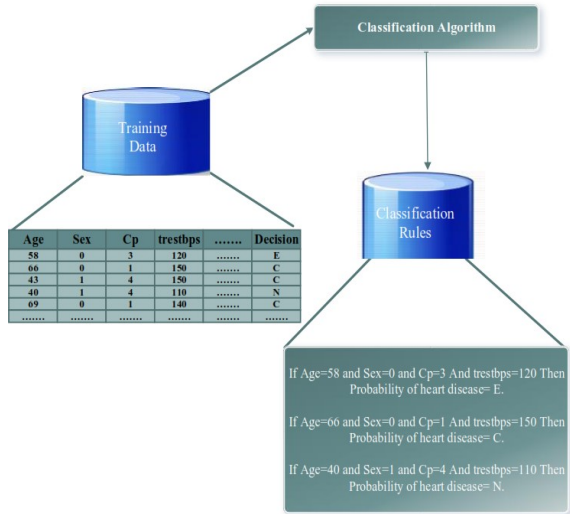

Fig.1: Building the classifier phase.

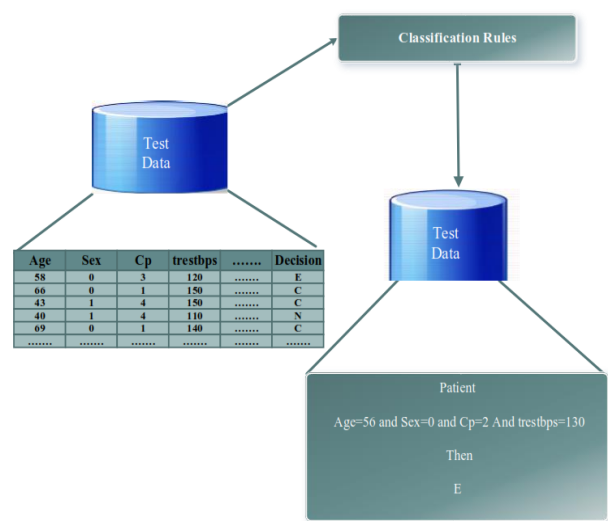

Fig.2: using classifier for classification

IV. CLASSIFICATION ALGORITHMS

This section presents an introduction to the classification algorithms that are used in our comparative study. As we have stated in the introduction, the purpose of the study is to investigate the stability and performance of the classification algorithms for predicting patient status from previous known patients data, using heart dieses related data. The employed classification algorithms are Decisions Tree, Naive Bayes, Discriminant, Random forest, and SVM.

A. *Decisions Tree*

A decision tree is considered to be one of the most popular approaches for producing classifiers [3]. It is equivalent to the flowchart structure in which internal node denotes as condition on an attribute, every branch represents the outcome of the condition and every leaf node represents the class label, and then the decision is taken after computing all attributes. A path from a root to a leaf represents classification rules [30]. In medical field decision trees determine the sequence of attributes. First, it produces a set of solved cases. Then the whole set is divided into training set and testing set. Where a training set is used for the induction of a decision tree. While the testing set is used to find the accuracy of an obtained solution [31].

B. *Naïve Bayes*

A naive Bayes classifier is a simple and efficient probabilistic classifier based on applying Bayes' theorem with strong (naive) independence assumptions [30]. It considers that the presence/absence of a particular attribute of a class is unrelated to the presence/absence of any other attribute when the class variable is given [5]. It helps to make computation process very easy and it has got better speed and accuracy for huge data [3]. Also, it have the ability of calculating the most possible output based on the input, and to add new raw data at run time [5].

*C. Discriminant*

Discriminant classifier works under the assumption that different classes generate data based on different Gaussian distributions. In the training stage the Gaussian distribution parameters for every class are estimated by the fitting function and in order to predict the class labels of new data, the trained classifier finds the class with the smallest misclassification cost [32].

*D. Random Forest*

It is an ensemble learning method for classification that is built by constructing a multiple decision trees at training time, and producing the class by voting of individual trees. It is Like Decision Tree, but the algorithm constructs a forest of decision trees with locations of attributes chosen at random. It has got the advantage of computing efficiency, improving the prediction accuracy without significantly increase of computational cost. It can also predict up to thousands of explanatory variables [30].

*E. Support Vector Machine (SVM)*

SVM has become more popular tool for machine learning tasks. It is supervised learning model that is applied mainly for classification, but it can also work for regression problems. The basic SVM works as binary classifier where the training data is divided into two classes. For multiclass problem, the basic SVM algorithm is executed repeatedly on the training data [31]. The SVM algorithm mapped feature vector into a higher dimensional vector space, where a maximum margin hyper-plane is established in this space. The distance from the hyper-plane to the nearest data point on each side is maximized. Maximizing the margin and thereby producing the largest possible distance between the separating hyper-plane and the instances on each side of it has been proven to reduce an upper bound on the expected generalization error.

## V. RESEARCH METHODOLOGY

The current research intends to improve the diagnosing of heart diseases by examining the patient's symptoms using data mining classification methods. To achieve this goal, a literature review was carried out to review the data mining works related to diagnose heart diseases. After that, five classifiers (e.g., Naïve Bayes, Decision Tree, Discriminant, Random Forest, and Support Vector Machine) were selected to build various classifiers. Each technique has different way to build classifier, which ensures that these techniques behave differently and produce different results. To underscore the practical viability of our approach, the selected classifiers have been implemented using MATLAB®. In all conducted experiments, the leave one out cross validation (holdout) model was adopted to evaluate algorithms' performance where original datasets are partitioned into different training and testing sets. In each validation cycle, one observation is hold out whereas the remaining observations act as training set and they are used to build classifier model. This step is repeated until all observations acted as test observations. Accordingly, the performance errors are calculated in each iteration, and the average performance is calculated at the end of validation.

*A.   Datasets description*

Two datasets were used in this study. The first one was obtained from the Cleveland Clinic Foundation [33], and It consists of 303 records, 297 are complete and 6 with missing/unknown values. The second one was Statlog dataset that is available at [34], and it consists of 270 completed records. Originally, both of these datasets have 76 attributes, but they were pre-processed to produce 14 attributes in an effort to reduce the number of variables. Consequently, we used these specific attributes (listed in Table 1) to allow comparison with other literatures.

TABLE 1

HEART FAILURE DATA SET WITH 14 ATTRIBUTES.

| # | Attribute Name | Attribute Information |
|---|---|---|
| 1 | #3  (age) | Age of the patient in years. |
| 2 | #4  (sex) | Represented as a binary number.<br>1 = male.<br>0 = female. |
| 3 | #9  (cp) | Chest pain type. Values range from 1 to 4.<br>Value 1: typical angina.<br>Value 2: atypical angina.<br>Value 3: non-anginal pain.<br>Value 4: asymptomatic. |
| 4 | #10 (trestbps) | Resting blood pressure measured in mm Hg on admission to the hospital. |
| 5 | #12 (chol) | Serum cholesterol of the patient measured in mg/dl. |
| 6 | #16 (fbs) | Fasting blood sugar of the patient. If greater than 120 mg/dl the attribute value is 1 (true), else the attribute value is 0 (false).<br>Value 1 = true.<br>Value 0 = false. |
| 7 | #19 (restecg) | Resting electrocardiographic results for the patient. This attribute can take 3 integer values 0, 1, or 2.<br>Value 0: normal.<br>Value 1: having ST-T wave abnormality (T wave inversions and/or ST elevation or depression of > 0.05 mV).<br>Value 2: showing probable or definite left ventricular hypertrophy by Estes' criteria. |
| 8 | #32 (thalach) | Maximum heart rate achieved of the patient. |
| 9 | #38 (exang) | Exercise induced angina. Values can be 0 or 1.<br>Value 1 = yes.<br>Value 0 = no. |
| 10 | #40 (oldpeak) | ST depression induced by exercise relative to rest. |
| 11 | #41 (slope) | Measure of slope for peak exercise. Values can be 1, 2, or 3.<br>Value 1: up sloping.<br>Value 2: flat.<br>Value 3: down sloping. |

| 12 | #44 (ca) | Number of major vessels (0-3) colored by fluoroscopy. Attribute values can be 0 to 3. |
|---|---|---|
| 13 | #51 (thal) | Represents heart rate of the patient. It can take values 3, 6, or 7. Value 3 = normal. Value 6 = fixed defect. Value 7 = reversable defect. |
| 14 | #58 (num) | Contains a numeric value between 0 and 4. Each value represents a heart disease or absence of all of them. Value 0     : < 50% diameter narrowing. (Absence of heart disease). Value 1 to 4: > 50% diameter narrowing. (Presence of different heart diseases). |

## B. Data Pre-Processing

The used data sets were collected from the University of California Irvine data repository. The selected data was checked for noise, inconsistency and missing values using distribution frequency while outlier detection was done using box plots. Noises and inconsistencies identified in the dataset were corrected manually, while missing values were replaced with the most similar value determined by Nearest Neighbor algorithm. Missing values are extremely common in medical records. However, these values must be treated before being used as they may lead to failure classification or incorrect disease prediction [35]. In fact, two common techniques could be applied for handling the datasets: deletion and imputation. Deletion is the most commonly proposed option for dealing with missing values where cases that cannot be processed are removed of. Despite the simple implantation of this method however, it is often considered unethical, especially when medical datasets are concerned. Furthermore, it may remove useful information. As an alternative, imputation method is used where missing values are replaced with estimated values.

In our research, the second approach is used. Specifically k-NN method has been used where a case is imputed using values from the k most similar cases. Briefly, to classify an unknown instance, the k-NN classifier calculates the distances between the point and points in the training data set using the Euclidean distance function [36] and then replaces it with the missing value. This has accomplished significant contribution from several medical domain researchers.

## C. Performance measures

Five common performance measures have been used to evaluate the accuracy of classification algorithms. These measure were selected because they are widely used to assess performance of classification models. The firs measure is Recall as shown in Equation 1, which measures how well a binary classification test correctly identifies a condition probability of correctly labeling members of the target class. The second measure is the Precision as shown in Equation 2, which measures probability that a positive prediction is correct. The third measure is accuracy as shown in Equation 3, which measures the performance of classification. The fourth measure is Specificity as shown in Equation 4, which measures how well a binary classification test correctly identifies the negative cases. The last measure is F-measure as shown in Equation 5, which measures probability that a positive prediction is correct [37].

$$Recall = \frac{t_p}{t_p + f_n} \tag{1}$$

$$Precision = \frac{t_p}{t_p + f_p} \tag{2}$$

$$Accuracy = \frac{t_p + t_n}{t_p + t_n + f_p + f_n} \tag{3}$$

$$sepcificity = \frac{t_n}{t_n + f_p} \tag{4}$$

$$F - measure = \frac{2 \times precision \times recall}{precision + recall} \tag{5}$$

Where:

tp is number of true positive.

tn is the number of true negative.

fp is the number of false positive

fn is the number of false negative

## VI. Results

In traditional healthcare systems, doctors relay on the signs or symptoms of patients to diagnosis heart diseases. However, many signs and symptoms are nonspecific, which may lead to unwanted errors and thus affecting the quality of provided health services. To overcome this issue, expert systems were developed to emulate health decision making with improved accuracy. This involves the correlation of various pieces of patients' information followed by obtaining specific patterns. In this research, we have applied five classification algorithms for predicting heart diseases based on pre-selected attributes.

For each dataset, we run the five classification algorithms using leave one out cross validation. The obtained results are presented in Tables 2 and 3. The results in Table 2 show that all classification algorithms are predictive and can give relatively correct answer about patient status. Notably, the recall measure shows that all classification model can work well with Cleveland dataset, but unfortunately it cannot distinguish which classification algorithm produces the most accurate results. Similarly, the specificity measure cannot tell us which classification algorithm produce the superior results. However, the remaining performance measures (i.e. accuracy, specificity and F-measure) have the same tendency which confirms that the decision tree is most accurate classifier among all classification algorithms.

TABLE 2

PERFORMANCE MEASURES USING CLEVELAND DATASET.

| Performance Measures | Decision Tree | Naïve Bays | Discriminant | Random Forest | SVM |
|---|---|---|---|---|---|
| Accuracy | 0.9901 | 0.7888 | 0.8350 | 0.9340 | 0.7657 |
| Specificity | 0.0000 | 0.0000 | 0.0000 | 0.0000 | 0.0000 |
| Precision | 0.9901 | 0.7888 | 0.8350 | 0.9340 | 0.7657 |
| Recall | 1.0000 | 1.0000 | 1.0000 | 1.0000 | 1.0000 |
| F-measure | 0.9950 | 0.8819 | 0.9101 | 0.9659 | 0.8673 |

Similarly, we run the five algorithms on Statlog dataset. The results obtained for this dataset are quite similar to those obtained for Cleveland dataset. The decision tree is the most accurate algorithm confirmed by the results of accuracy, precision and F-measure. Both Recall and Specificity measure cannot tell us any useful feedback because they cannot distinguish among algorithms. The notable observation in both datasets is that the SVM cannot beat other algorithm even though this algorithm is considered one of top classification algorithm for health and medical datasets.

Another important issue that should be emphasized in this study is the ranking stability of classification algorithm over both datasets and across multiple performance measures. Ranking stability of algorithm means that this algorithm should always produce accurate results and rank high over all datasets and across all evaluation measures. From our results, we can observe that the decision tree was ranked first across all evaluation measures and in both datasets, followed by Random forest in the second position. This can give us an important conclusion about the role of decision rules in classifying data. Both decision tree and Random forest have relatively same mechanism but the Random forest can build ensemble of decision tree. Although ensemble learning has been proved to produce superior results, but in our case the decision tree has outperformed its ensemble version. Finally, from the above results we can figure out that the Naïve Bayes and Discriminant algorithm produce relatively similar results.

TABLE 3

PERFORMANCE MEASURES USING STATLOG DATASET.

| Performance Measures | Decision Tree | Naïve Bays | Discriminant | Random Forest | SVM |
|---|---|---|---|---|---|
| Accuracy | 0.9815 | 0.8037 | 0.8259 | 0.9148 | 0.7556 |
| Specificity | 0.0000 | 0.0000 | 0.0000 | 0.0000 | 0.0000 |
| Precision | 0.9815 | 0.8037 | 0.8259 | 0.9148 | 0.7556 |
| Recall | 1.0000 | 1.0000 | 1.0000 | 1.0000 | 1.0000 |
| F-measure | 0.9907 | 0.8912 | 0.9047 | 0.9556 | 0.8608 |

## VII. CONCLUSION AND FUTURE WORK

In this research, different data mining techniques were studied to enhance the early detection of heart diseases. Specifically, five classifiers were utilized and implemented using MATLAB tool to emulate health decision making with improved accuracy. Moreover, an extensive analysis was conducted using two datasets. For each dataset, the selected classifiers were run using leave one out cross validation. Results showed that all classification algorithms are predictive and can give relatively correct answer. However, it confirms that the decision tree was ranked first across all evaluation measures and in both datasets, followed by Random forest in the second position. This can give us an important conclusion about the role of decision rules in classifying data. Both decision tree and Random forest have relatively same mechanism but the Random forest can build ensemble of decision tree.

As a future work, we will use the research described here as a foundation for the development of effective prediction system to enhance medical care and reduce costs. That being said, this is a large topic and there are numerous opportunities for additional research that would significantly extend the functionality of the current research. For example: Considering other kinds of diseases that are biologically related to heart diseases. It would also possible to plug in the prediction system into other systems such as a context-aware access control security system to support the security fundamentals of healthcare systems.

## ACKNOWLEDGMENT

The authors are grateful to the Applied Science Private University, Amman-Jordan, for the full financial support granted to cover the publication fee of this research article.